\documentclass[aps,prl,twocolumn,superscriptaddress,preprintnumbers]{revtex4}

\usepackage{epsf,epsfig,amsmath,amssymb,amsfonts}
\usepackage{color}
\usepackage{slashed}
\usepackage{hyperref}

\newsavebox{\ns}
\newsavebox{\dbrane}

\def\be{\begin{equation}}
\def\ee{\end{equation}}
\def\bea{\begin{eqnarray}}
\def\eea{\end{eqnarray}}

\def\Dslash{\,\,{\raise.15ex\hbox{/}\mkern-12mu D}}
\def\Dbarslash{\,\,{\raise.15ex\hbox{/}\mkern-12mu {\bar D}}}
\def\delslash{\,\,{\raise.15ex\hbox{/}\mkern-9mu \partial}}
\def\delbarslash{\,\,{\raise.15ex\hbox{/}\mkern-9mu {\bar\partial}}}
\def\pslash{\,\,{\raise.15ex\hbox{/}\mkern-9mu p}}
\def\calDslash{\,\,{\raise.15ex\hbox{/}\mkern-12mu {\cal D}}}

\newcommand\diff{\mbox{d}}

\newcommand{\nn}{\nonumber \\}

\newcommand{\dd}{\diff}

\allowdisplaybreaks

\begin{document}

\title{Classical Yang-Baxter Equation from Supergravity}

\preprint{APCTP Pre2017-017}         

\preprint{IPM/P-2017/051}
\vskip 1 cm

 \author{I. Bakhmatov}
 \affiliation{Asia Pacific Center for Theoretical Physics, Postech, Pohang 37673, Korea}
 \affiliation{Institute of Physics, Kazan Federal
University, Kremlevskaya 16a, 420111, Kazan, Russia}
 \author{\"O. Kelekci}
 \affiliation{Faculty of Engineering, University of Turkish Aeronautical Association, 06790 Ankara, Turkey}
 \author{E. \'O Colg\'ain}
 \affiliation{Asia Pacific Center for Theoretical Physics, Postech, Pohang 37673, Korea}
\author{M. M. Sheikh-Jabbari}
\affiliation{School of Physics, Institute for Research in Fundamental Sciences (IPM), P.O.Box 19395-5531, Tehran, Iran}

\begin{abstract}

\noindent 
We promote the open-closed string map, originally formulated by Seiberg \& Witten, to a solution generating prescription in generalized supergravity. The approach hinges on a knowledge of an antisymmetric bivector $\Theta$, built from antisymmetric products of Killing vectors, which is specified by the equations of motion. In the cases we study, the equations of motion reproduce the Classical Yang-Baxter equation (CYBE) and $\Theta$ is the most general $r$-matrix solution. Our work generalizes Yang-Baxter deformations to non-coset spaces and unlocks gravity as a means to classify $r$-matrix solutions to the CYBE.  

\end{abstract}

\maketitle

\setcounter{equation}{0}

\section{Introduction} \label{Introduction}

Generating exact solutions to gravity theories is a fine, but well-practiced art \cite{Ehlers, Geroch:1970nt}.  In this regard, supergravity theories, being consistent backgrounds of string theory, are especially rich. These theories often inherit symmetries of the parent theory, including T-duality \cite{Buscher:1987sk, Buscher:1987qj}, which is well-known \cite{Bakas:1994np, Bakas:1994ba} to masquerade as classic solution generating techniques \cite{Ehlers, Geroch:1970nt}. In the presence of anomalies \cite{Gasperini:1993nz, Alvarez:1994np,Elitzur:1994ri}, this aspect of T-duality, including its generalizations \cite{delaOssa:1992vci, Giveon:1993ai, Klimcik:1995ux, Shelton:2005cf}, is obscured. In recent years, driven by developments in integrable deformations of $\sigma$-models \cite{Klimcik:2002zj, Klimcik:2008eq, Delduc:2013qra, Kawaguchi:2014qwa}, especially \cite{Arutyunov:2013ega, Arutyunov:2015qva}, we have started to understand these anomalies through a modification of supergravity, called  ``generalized supergravity" \cite{Arutyunov:2015mqj} (also \cite{Wulff:2016tju}). The modification is encoded in an extra Killing vector $I$, with usual supergravity recovered when $I =0$. Exotic though it may seem, from the perspective of lower dimensions, this theory is no more than matter-coupled Einstein gravity.

In this letter, we promote the closed string to open string map of Seiberg \& Witten \cite{Seiberg:1999vs}, or more accurately, the inverse map, to a simple, effective solution generating technique. This map was initially introduced in \cite{Seiberg:1999vs}, where it was argued that open strings attached to D-branes in a constant $B$-field probe a noncommutative (NC) space, whose metric is the open string metric. It is in fact fairly ubiquitous, applicable even for non-constant $B$-field. Its connection to T-duality has been exploited in actions that make non-geometric fluxes manifest \cite{Andriot:2011uh, Andriot:2012wx} and string theory explanations \cite{Hellerman:2011mv, Hellerman:2012zf} of the $\Omega$-deformation \cite{Moore:1997dj, Nekrasov:2002qd}. More recently, it was noted \cite{Araujo:2017jkb, Araujo:2017jap} that the closed-open string map undoes integrable deformations of $\sigma$-models \cite{Klimcik:2002zj, Klimcik:2008eq, Delduc:2013qra, Kawaguchi:2014qwa}. 

Building on the open-closed string map, we provide a solution generating prescription that is accessible to the gravity community. 
Starting from a supergravity solution with metric $G$ and zero NSNS two-form, or $B$-field, one turns on an antisymmetric bivector $\Theta$. This then defines ``open string data", which upon inverting a single matrix, generates ``closed string data", namely a new metric $g$ and $B$-field. The transformed dilaton (up to a constant shift) is determined from a well-known T-duality invariant, while the Killing vector $I$ is simply the divergence of $\Theta$ \cite{ Araujo:2017enj}. Together, ($g, B, \Phi, I$) present a consistent NS sector of generalized supergravity. For the RR sector, field strengths are determined from the non-zero Page forms \cite{Page:1984qv, Marolf:2000cb}, which are the open string counterparts of the RR fields. In turn, the lower-dimensional forms are specified by a descent procedure through contracting $\Theta$, and hence the new solution is completely determined by the bivector.

Concretely, we propose that $\Theta$ is a linear combination of antisymmetric products of Killing vectors of the original geometry with constant coefficients, where the relation between constants is in turn fixed by the the equations of motion (EOMs) of generalized supergravity. To show the workings, we consider $AdS_2 \times S^2$ and Schwarzschild spacetimes, respectively a coset and non-coset space. Remarkably, the algebraic conditions on the constants are none other than the Classical Yang-Baxter equation (CYBE) associated with the isometry group of the original solution and $\Theta$ is the most general $r$-matrix solution to the CYBE. In short, the generalized supergravity EOMs yield the CYBE. 


We recall that the CYBE arises in the classical limit of the Yang-Baxter equation, which is a hallmark of integrability, or exact solvability, in statistical mechanics, quantum field theory, differential equations, knot theory, quantum groups, etc \cite{Jimbo, Perk:2006wiu}. Of special interest, $r$-matrix solutions to the CYBE are related to Drinfel'd twists \cite{Drinfeld:1985rx} in NC field theory \cite{Chaichian:2004za}. Through this work, we provide the first example of a gravitational set-up with an innate knowledge of the CYBE.  

\section{Prescription}
Here we give a prescription for generating new (generalized) supergravity solutions from existing solutions with zero $B$-field. Our methodology will ultimately be justified by the end result. We start by describing the NS sector transformation, before addressing the complementary RR sector. We focus on IIB supergravity. 

\subsection{NS sector}
We recall the open-closed string map of Seiberg \& Witten \cite{Seiberg:1999vs}, which we recast in the following form:
\be
\label{matrix}
(g + B)_{\mu \nu} = ( G^{\mu \nu} + \Theta^{\mu \nu})^{-1},  
\ee
where $(g, B)$, $(G, \Theta)$ are respectively closed string and open string fields. The metrics $g, G$ are of course symmetric, $B, \Theta$ are their antisymmetric counterparts and $B = 0$ implies $\Theta = 0$, and vice versa. Our approach is to interpret the metric of the original solution as the open string metric $G$, add a deformation parameter $\Theta$, then generate a new metric $g$ and $B$-field. This map works for generic supergravity solutions, not necessarily coset spaces, for example the Schwarzschild solution. For spacetimes with $U(1)^2$ isometry our method reduces to T-duality shift T-duality (TsT) transformations \cite{Lunin:2005jy}, but it is more generally applicable. 

The NS sector of supergravity comprises, in addition to $g$ and $B$, a scalar dilaton $\Phi$. Moreover, when further extended to generalized supergravity, one encapsulates the modification of usual supergravity in a single one-form $X$ \cite{Arutyunov:2015mqj}: 
\be
\label{X}
X \equiv \dd \Phi + i_{I} B + I, 
\ee
where $I$ is the one-form related to the Killing vector; setting $I = 0$, we recover usual supergravity. The NS sector of generalized supergravity is hence characterized by $(g, B, \Phi, I)$. We note that the $B$-field is specified up to the $\Lambda$-gauge transformation, $B\to B+\dd \Lambda$ and  (\ref{matrix}) and (\ref{X}) are written in a particular $\Lambda$-gauge, while $X$, which appears in the EOMs of generalized supergravity,  is $\Lambda$-gauge invariant \cite{Arutyunov:2015mqj, Araujo:2017jap}. However, this leaves the residual symmetry of shifting $\Phi$ by a constant, $\Phi \rightarrow \Phi + \Phi_0$, without changing $B$. 

Having specified how $g, B$ are generated, we turn our attention to $\Phi$ and $I$.  The dilaton transformation follows from the observation that there is a well-known T-duality invariant \cite{Buscher:1987sk, Buscher:1987qj}, 
\be
\label{dilaton}
e^{-2 \delta \Phi} \sqrt{\det g_{\mu\nu}} = \sqrt{\det G_{\mu\nu}}, 
\ee
where $\delta \Phi$ is the dilaton shift, modulo the constant $\Phi_0$. This ensures our prescription is not at odds with TsT. Lastly, the status of the final solution, being either supergravity or generalized supergravity, may be read off from the divergence of $\Theta$ \textit{with respect to the original metric} \cite{Araujo:2017enj}, 
\be
\label{I_theta}
I^{\mu} = \nabla_{\nu} \Theta^{\nu \mu}.  
\ee
The origin of this equation can be explained in terms of a consistency condition arising from the $\Lambda$-gauge invariance of D-brane actions \cite{Araujo:2017enj}. For TsT transformations, $\Theta$ is a constant \cite{Araujo:2017jap}, so that $I = 0$ and the final solution is a \textit{bona fide} supergravity solution.  This completes our description of the NS sector transformation. 

\subsection{RR sector}
We turn attention to the RR sector. The standard treatment in T-duality, or any frame change, is that there is a Lorentz transformation acting on a bispinor constructed from the RR field strengths $F_n$ \cite{Hassan:1999bv, Sfetsos:2010uq, Borsato:2016ose}. 
Here we adopt a novel approach, which makes the role of $\Theta$ manifest. 

We recall the Page forms \cite{Page:1984qv, Marolf:2000cb}  \footnote{We employ the notation $B^2 = B \wedge B$, etc., and later $A \lrcorner B = \frac{1}{p!} A^{\mu_1\dots \mu_p} B_{\mu_1 \dots \mu_p \nu_{p+1} \dots \nu_{q} }$ for $p$-form $A$ and $q$-form $B$ with $q  \geq p$.}, 
\be
\label{Page_forms}
\begin{aligned}
Q_1 &= F_1, \quad Q_3 = F_3 + B F_1,\\ Q_5 &= F_5 + B F_3 + \frac{1}{2} B^2 F_1, \\
Q_7 &= - {*}F_3 + B F_5 + \frac{1}{2} B^2 F_3 + \frac{1}{3!} B^3 F_1, \\
Q_9 &= {*}F_1 - B\, {*}F_3 + \frac{1}{2} B^2 F_5 + \frac{1}{3!} B^3 F_3 + \frac{1}{4!} B^4 F_1, 
\end{aligned}
\ee
which may be viewed as the completion of the open-closed string map (\ref{matrix}) to the RR sector. It was first noted in \cite{Araujo:2017enj} that the EOMs of generalized supergravity simplify when expressed as Page forms : 
\be
\label{Q_EOM}
\dd Q_{2 n-1} = i_{I}  Q_{2 n+1}, \quad n = 1, 2, 3, 4. 
\ee 

It is well-known that the Page charges, integrals of Page forms over compact cycles, can be quantized \cite{Marolf:2000cb}. In particular, in AdS/CFT the quantized charges correspond to ranks of the gauge groups in the dual gauge theory. Now, recall that $\Theta$ is a deformation parameter, which we can continuously set to zero. Since the Page charges can only change discretely, they can not depend on $\Theta$, leading to the conclusion that they are invariant. Since the cycles do not change, this implies that the corresponding Page forms are indeed invariant. This invariance of the non-zero Page forms constitutes the basis of a consistent treatment of the RR sector. 

In our proposal, the initial open string data is completed by specifying the Page forms. Since originally the $B$-field was absent, this implies there exist given Page forms in the new solution that satisfy
\be
Q_{2 n+1} = F_{2 n+1}, 
\ee
where $n$ is determined by the original non-zero field strengths $F_{2 n +1}$. The remaining Page forms are generated through descent by contracting $\Theta$, 
\be
\label{descent}
Q_{2 (n-p) +1} = \frac{1}{p!} \Theta^p \lrcorner  Q_{2 n+1}. 
\ee
Given (\ref{I_theta}), for each $p$, (\ref{descent}) is a solution to (\ref{Q_EOM}). Unraveling the Page forms using the generated $B$-field, one arrives at the final expression for the RR field strengths. It is worth emphasizing again that all information about the deformation is encoded in $\Theta$. 

Putting the NS and RR sectors together, one finds a prescription for writing down the deformed geometry solely on the basis of a knowledge of $\Theta$. Let us recapitulate the key steps:
\begin{enumerate}
\item Invert matrix $G^{-1} + \Theta$ to determine $g, B$.

\item Calculate $\delta \Phi$ from a known T-duality invariant.

\item Determine $I$ from divergence of $\Theta$. 

\item Non-zero Page forms are invariant.

\item Determine the remaining Page forms via descent equation (\ref{descent}).
\end{enumerate} 

\section{Prescription at work}
Let us turn our attention to some examples, where we employ the above prescription and solve for $\Theta$. 

\subsection{ Example I: $AdS_2 \times S^2$}
Here we illustrate our prescription with the geometry $AdS_2 \times S^2 \times T^6$, which corresponds to the near-horizon of intersecting D3-branes. The initial supergravity solution is,  
\be\begin{aligned}
\dd s^2 &= \frac{(-\dd t^2 + \dd z^2)}{z^2} + \dd \zeta^2 + \sin^2 \zeta \dd \chi^2 + \dd s^2 (T^6), \\
F_5 &= ( 1+ *_{10})  \frac{1}{\sqrt{2} z^2} \dd t \wedge \dd z \wedge ( \omega_r - \omega_i),  
\end{aligned}\ee
where we define the three-forms $\omega_r - i \omega_i  = \Omega_3$, with $\Omega_3$ being the complex $(3,0)$-form on the torus. Observe that both the $B$-field and dilaton are zero. 

We consider the following ansatz for the deformation 
\be
\Theta^{tz} = \Theta_1( t, z), \quad \Theta^{\zeta \chi} = \Theta_2 ( \zeta, \chi). 
\ee
This ansatz honors the direct-product structure of the geometry, leaving us the task of solving for two functions. Note, we have not assumed that $\Theta$ is an antisymmetric product of Killing vectors from the outset, instead this is forced upon us by the EOMs as we now show. 

Following our recipe, we arrive at a new solution to generalized supergravity, which is fully determined modulo $\Theta_i, i = 1, 2$. To give a flavor of the output, suppressing the torus, we record $(g, B, \Phi)$, 
\be\begin{aligned}
\dd s^2 &= \frac{z^2 ( - \dd t^2 + \dd z^2 ) }{z^4 - \Theta_1^2} + \frac{\dd \zeta^2 + \sin^2 \zeta \dd \chi^2}{1 + \Theta_2^2 \sin^2 \zeta}  , \\
B &= \frac{\Theta_1}{z^4 - \Theta_1^2} \dd t \wedge \dd z - \frac{\Theta_2 \sin^2 \zeta}{1 + \Theta_2^2 \sin^2 \zeta} \dd \zeta \wedge \dd \chi, \\
e^{2 \Phi} &= \frac{e^{2 \Phi_0} z^4}{(z^4 - \Theta_1^2)(1 + \Theta_2^2 \sin^2 \zeta)}. 
\end{aligned}\ee
We have presented the complete solution in supplemental material. Note, $B$-field is pure gauge, so we could set it to zero, if desired. 

When solving for $\Theta_i$, it is most effective to initially recall that $I$ is Killing and solve the Killing equation, $\nabla_{\mu } I_{\nu} + \nabla_{\nu} I_{\mu} =0$. This determines $\Theta_i$ up to eight integration constants: 
\begin{align}
\label{theta1} \Theta_1 &= c_1 t z +  c_2 z ( t^2 -z^2) + c_3 z + c_4 z^2, \\
\label{theta2} \Theta_2 &= c_5 \cos \chi + c_6 \sin \chi + c_7 \cot \zeta + \frac{c_8}{\sin \zeta}.   
\end{align}
As will be clear soon, modulo the $c_4, c_8$ terms that are forced to vanish, $\Theta$ has already been determined as a linear combination of antisymmetric products of Killing vectors. 

We next study the Einstein equation, the EOM for the $B$-field and the dilaton EOM, where the first two equations lead to the same set of constraints: 
\begin{align}
\label{ads_constraint} \kappa^2 &= - c_1^2 + 4 c_2 c_3, \\ 
\label{s_constraint} \kappa^2 &=  c_5^2 + c_6^2 + c_7^2, \\ 
c_4 &= c_8 =0. 
\end{align}
Here we have redefined the constant shift in the dilaton $e^{2 \Phi_0} = 1 + \kappa^2$. Eq.\eqref{ads_constraint} recently appeared in \cite{Kyono:2017jtc}.
Note that these EOMs split between the $AdS_2$ and $S^2$ spaces, but are connected via the constant dilaton shift. The dilaton EOM is satisfied given \eqref{ads_constraint} and \eqref{s_constraint}.

One can also check the EOMs involving the RR field strengths. Our descent procedure for the Page forms ensures that the EOMs are satisfied by construction and hence one finds no further constraints. Therefore, subject to the constraints, we have the most general solution for $\Theta$. We now turn to the interpretation. 

Let us begin with $\kappa = 0$. Evidently, there is no real solution to (\ref{s_constraint}), so this precludes a deformation of the two-sphere. In contrast, there is an allowed deformation of the $AdS_2$ space with  two free parameters. We can interpret both of these results from the perspective of the associated homogeneous CYBE, provided $\Theta$ is an $r$-matrix solution.
To do so, let us label the six Killing vectors of the $AdS_2 \times S^2$ geometry as 
\be
\label{Killing}
\begin{aligned}
T_1 &= - t\, \partial_t - z\, \partial_z, \quad T_2 = - \partial_t, \\ 
T_3 &= - (t^2 + z^2) \,\partial_t - 2 t z \,\partial_z, \\
T_4  &= \partial_{\chi}, \quad T_5 =  - \cos \chi\, \partial_{\zeta} + \cot \zeta \sin \chi\, \partial_{\chi}, \\
T_6 &= \sin \chi\, \partial_{\zeta} + \cot \zeta \cos \chi\, \partial_{\chi}. 
\end{aligned}\ee
Constructing the most general $r$-matrix $r = \frac{1}{2} r^{ij} T_i \wedge T_j$, and substituting the components of the $r$-matrix into the homogeneous CYBE corresponding to the $\frak{sl}(2)\oplus \frak{su}(2)$ algebra, we arrive at the constraints: 
\be
r^{12} r^{31} = (r^{23})^2, \quad (r^{45})^2 + (r^{56})^2 + (r^{64})^2 = 0. 
\ee
Relabeling the components of the $r$-matrix, $r^{12} = -c_3$, $2 \, r^{23} = c_1$,  $r^{31} = -c_2$, $r^{45} = -c_5$, $r^{56} = - c_7$, $r^{64} = - c_6$, it is easy to check that the (non-zero) $r$-matrix is 
\be
r =  \Theta_1 \partial_t \wedge \partial_z.  
\ee
Thus, when $\kappa = 0$, (\ref{ads_constraint}) and (\ref{s_constraint}) are essentially the homogeneous CYBE for the Lie algebras $\frak{sl}(2)$ and $\frak{su}(2)$, respectively. It is well known that there is a redundancy in the CYBE and $r$-matrix solutions are equivalent up to  automorphisms. In the geometry, these correspond to coordinate changes and it is easy to check that under a special conformal transformation, one can set $c_1 = c_2 = 0$, while under a translation, one can set $c_1 = c_3 = 0$, leaving one parameter.

When $\kappa \neq 0$, namely when there is a constant dilaton shift, we find an apparent five-parameter class of deformations of $AdS_2 \times S^2$, the redundancy of which can be removed again by coordinate change leaving a single parameter $\kappa$. This corresponds to a solution to the modified CYBE with modification $\kappa$. Modulo a coordinate transformation, 
\be\begin{aligned}
\rho &= \frac{z^2 -1 -t^2}{2 z}, \quad \cos \tilde{t} = \frac{z^2 + 1 -t^2}{\sqrt{4 t^2 + (z^2 + 1 -t^2)^2}}, \\
r &= \cos \zeta, \quad \varphi = \chi, 
\end{aligned}\ee
where we have added tildes to differentiate new coordinates, we can recover the known solution in the literature \cite{Arutyunov:2015mqj} through the choice $c_2 = - \frac{\kappa}{2}$, $c_3 = -\frac{\kappa}{2}$, $c_7 = - \kappa$. 

The important take-home lesson from this simple example is that the CYBE naturally emerges from the open-closed string map and the EOMs of generalized supergravity. Once the CYBE is imposed, we are guaranteed a new supergravity solution where $\Theta$ corresponds to the $r$-matrix. The well-known redundancy of the r-matrix under automorphisms corresponds to coordinate changes in the geometry.

\subsection{Example II: Schwarzschild} 
To confirm that the previous analysis was no fluke, we repeat for another geometry, the Schwarzschild  black hole,  
\be
\dd s^2 = - \left( 1- \frac{2 m}{r} \right) \dd t^2 + \frac{\dd r^2}{(1 - \frac{2 m}{r} )} + r^2 \left( \dd \zeta^2 + \sin^2 \zeta \dd \chi^2 \right). \nonumber
\ee
In contrast to the previous example, Schwarzschild does not admit a coset construction and is unlikely to be an integrable $\sigma$-model background. Furthermore, for this example it is difficult to solve for $\Theta$ directly, so we choose $\Theta$ to an antisymmetric product of Killing vectors:
\be
\label{theta_schwarzschild}
\begin{aligned}
\Theta^{t \zeta} &= - \epsilon \cos \chi + \lambda \sin \chi, \\
\Theta^{t \chi} &= \delta + \cot \zeta \left( \epsilon \sin \chi + \lambda \cos \chi \right), \\
\Theta^{\zeta \chi} &= \alpha \cos \chi - \beta \cot \zeta + \gamma \sin \chi. 
\end{aligned}
\ee
Note, this corresponds to 
\be
\begin{aligned}
\Theta &=& \alpha \, T_4 \wedge T_5 + \beta \, T_5 \wedge T_6 + \gamma \, T_6 \wedge T_4 \\
&+& \delta \, T_2 \wedge T_4 + \epsilon \, T_2 \wedge T_5 + \lambda \, T_2 \wedge T_6,  
\end{aligned}
\ee
where $\alpha, \beta$, etc are constants and we have employed (\ref{Killing}). While this is ostensibly the same form as the $r$-matrix, an important distinction is that the coefficients are not fixed. Before proceeding, we remark that the original geometry is Ricci-flat with no RR sector.

As an initial consistency check on $\Theta$, one confirms from (\ref{I_theta}) that $I = \beta \, T_4 + \gamma \, T_5 + \alpha \, T_6$ is a valid Killing vector. We now repeat the same matrix inversion from the open-closed string map and substitute into the EOMs. Assuming non-zero coefficients, the EOMs are satisfied provided,  
\be
\begin{aligned}
0 &= \beta \epsilon - \delta \gamma = \alpha \epsilon - \gamma \lambda = \alpha \delta - \lambda \beta, \\
0 &= \alpha^2 + \beta^2 + \gamma^2. 
\end{aligned}
\ee
The key observation now is that these equations are the same as the homogenous CYBE for the Lie algebra $\frak{u}(1) \oplus \frak{su}(2)$, in line with our expectations. It is worth stressing that our statement supergravity recovers the CYBE holds beyond strict coset geometries.

Indeed, the CYBE for this algebra involves selecting three generators from four, so we get precisely four equations, only three of which are independent. Here, without an RR sector, the constant shift in the dilaton makes no difference, so we cannot consider the modified CYBE. It is easy to see that $\alpha = \beta = \gamma = 0$, so the only permitted deformation involves $\delta, \epsilon, \lambda$ with no constraint. However, here again we encounter a redundancy and two of these parameters can be removed using two-sphere rotations.

\section{Discussion}
Let us review what has been achieved. Our main result is providing a prescription through which the CYBE emerges from the EOMs of a gravity theory, thus reducing the task of identifying $r$-matrix solutions to the CYBE to solving generalized supergravity EOMs. More precisely, starting from a given supergravity solution, with zero $B$-field, we have promoted the open-closed string map to a solution generating prescription. The solution is completely specified by a bivector $\Theta$, determined by the EOMs, and a knowledge of it is enough to simply write down the resulting solution. Our prescription for the RR sector employs a simple descent procedure, where lower-dimensional Page forms are induced. 

Our proposal follows from attempts to decipher the Yang-Baxter $\sigma$-model \cite{Klimcik:2002zj, Klimcik:2008eq, Delduc:2013qra, Kawaguchi:2014qwa}, simplify it and make it accessible. However, it goes beyond Yang-Baxter $\sigma$-models. As advertised, having adopted gravity as our medium, we are no longer shackled to cosets.  One can now experiment with new geometries, in the process generating large classes of exotic solutions. Secondly, we do not assume integrability via an $r$-matrix solution to the CYBE, but taking a step back to see the wood from the trees, we observe that the CYBE emerges. Thirdly, we note that one can easily derive rich solutions to the modified, versus homogeneous CYBE, through a constant dilaton shift. This itself is a residual symmetry left over in the field $X$ after the $\Lambda$-gauge is fixed \footnote{Please contrast with \cite{Arutyunov:2015mqj}, where a vanishing dilaton necessitates an ad hoc rescaling of the RR sector.}. 

Without assuming $\Theta$ to be an antisymmetric product of Killing vectors, we have solved the EOMs directly for deformations of $AdS_2 \times S^2$ to confirm that this must be the case. We strongly suspect that given any initial solution with isometries, then $\Theta$ is always an antisymmetric product of Killing vectors related to an $r$-matrix solution to the CYBE of the associated Lie algebra. While it would be intriguing to identify counterexamples, the fact that the algebraic CYBE can emerge from the dynamical EOMs of a gravity theory is striking. Bearing in mind that the classification of $r$-matrix solutions to the CYBE becomes arduous  as the algebra becomes larger \cite{Stolin1, Stolin2, Stolin3}, gravity offers a seemingly simple alternative. Furthermore, it would be interesting to understand the relation between integrability and the CYBE, since as we have seen with the Schwarzschild solution, the CYBE emerges, whether integrability is present or not.

Finally, based on intuition gained from several examples, we conjecture that $\Theta^{[\alpha \rho}\nabla_\rho\Theta^{\beta \gamma]}=0$ is a consistency condition arising from the generalized supergravity EOMs. This condition may be viewed as the Jacobi identity for an algebra of coordinates on a noncommutative, but associative space, $[X^\mu,X^\nu]=i\Theta^{\mu\nu}(X)$. If this conjecture holds, then the open string frame is more than just a name and the system indeed describes open strings with noncommuting endpoints. This has far-reaching implications for AdS/CFT. The same Jacobi identity also appears in the vanishing of $R$-flux in the context of Double Field Theory \cite{Hull:2009mi, Siegel:1993xq, Siegel:1993th}, where it ensures a geometric description \cite{Andriot:2011uh, Andriot:2012wx, Hassler:2016srl, Sakamoto:2017cpu}. These connections between integrable models, supergravity and noncommutativity warrant further study. 

\section*{Acknowledgements}
We thank K. Sfetsos, D. Thompson, K. Yoshida for discussion and especially F. Hassler, Y. Lozano for comments on the final manuscript. I. B. is partially supported by the Russian Government program for the 
competitive growth of Kazan Federal University. 
M. M.~Sh-J. is supported in part by the Saramadan Iran  Federation, the junior research chair on black hole physics by the Iranian NSF 
and the ICTP network project NT-04. O. K. would like to thank Nesin Mathematics Village (Izmir, Turkey) for hospitality, where part of this work was done. E. \'O C. thanks Kyoto University for hospitality during the workshop ``Noncommutative geometry, duality and quantum gravity", Sep 4-6 2017.

\end{document}